\documentclass[prl,twocolumn,showpacs,preprintnumbers,amsmath,amssymb]{revtex4-1}
\usepackage{amsmath}
\usepackage{epsfig}
\usepackage{graphics}
\begin{document}
\title{Decay of polarons and molecules in a strongly polarized Fermi gas}
\author{G.\ M.\  Bruun}
\affiliation{Mathematical Physics, Lund Institute of Technology, P.\ O.\ Box 118, SE-22100 Lund, Sweden}
\author{P.\ Massignan}
\affiliation{Grup de F\'isica Te\'orica, Universitat Aut\`onoma de Barcelona, 08193 Bellaterra, Spain }
\affiliation{ICFO-Institut de Ci\`encies Fot\`oniques, Mediterranean Technology Park, 08860 Castelldefels (Barcelona), Spain.}

\date{\today}
\begin{abstract}
The ground state of an impurity immersed in a  Fermi sea changes from a polaron to a molecule as the interaction strength is increased.
 We show here that the coupling between these two states is strongly suppressed due to a combination of phase space effects and Fermi statistics, and that it vanishes much faster than the energy difference between the two states, thereby confirming the first order nature of the polaron-molecule transition. In the regime where each state is metastable, we find quasiparticle lifetimes which are much longer than what is expected for a usual Fermi liquid. Our analysis indicates that the decay rates are sufficiently slow to be experimentally observable.
 \end{abstract}

\pacs{03.75.Ss, 
05.30.Fk, 
67.85.Lm 
}

\maketitle

The concept of a quasiparticle, developed by Landau in his theory of Fermi liquids~\cite{Landau}, is  fundamental  to our understanding of low-energy excitations in many-body systems~\cite{BaymPethick}. 
 It has been successfully applied to describe systems spanning many energy scales ranging from
 electrons in a metal to atomic nuclei and quark-gluon plasmas.
In a recent experiment on ultracold gases, quasiparticle physics has been analyzed in depth by studying the properties of a few impurities immersed in a Fermi sea (FS)~\cite{Schirotzek}.
 Theoretically, one has presently a good understanding of the weak coupling limit of this problem where the ground state is a screened impurity, a state which is referred to as a polaron, and of the strong coupling limit where the ground state consists of a molecule formed by the impurity and a particle in the FS~\cite{Prokofev,ChevyMora}. 
 In the intermediate regime, the ground state changes from a polaron to a molecule as the interaction is increased, but the nature of this 
 transition which involves a reorganization of the FS is  not clear at present. The polaron-molecule transition  is closely related to the problem of 
mixtures of dilute protons in neutrons at subnuclear density, which are encountered for example in stellar collapse. 
The protons may form either polarons or two-body bound states (deuterons) with the majority neutrons, in analogy with what is considered in the present paper. 
 

We address here the polaron-molecule transition using a diagrammatic expansion in the number of holes in the FS. The coupling between the polaron and the molecule is shown to scale with their energy difference $\Delta\omega $ as $|\Delta\omega|^{9/2}$. 
 The high power is a result of phase space effects and Fermi statistics.  It should be compared with the usual $\Delta\omega^2$ 
  scaling of the quasiparticle damping rate in a Fermi liquid~\cite{BaymPethick}. This scaling implies long lifetimes and a first order transition between the polaron and the molecule. 
  Our analytical insights are confirmed by numerical calculations, and in the conclusions we discuss how the decay can be measured.

\begin{figure}
\begin{center}
	\includegraphics[width=\columnwidth]{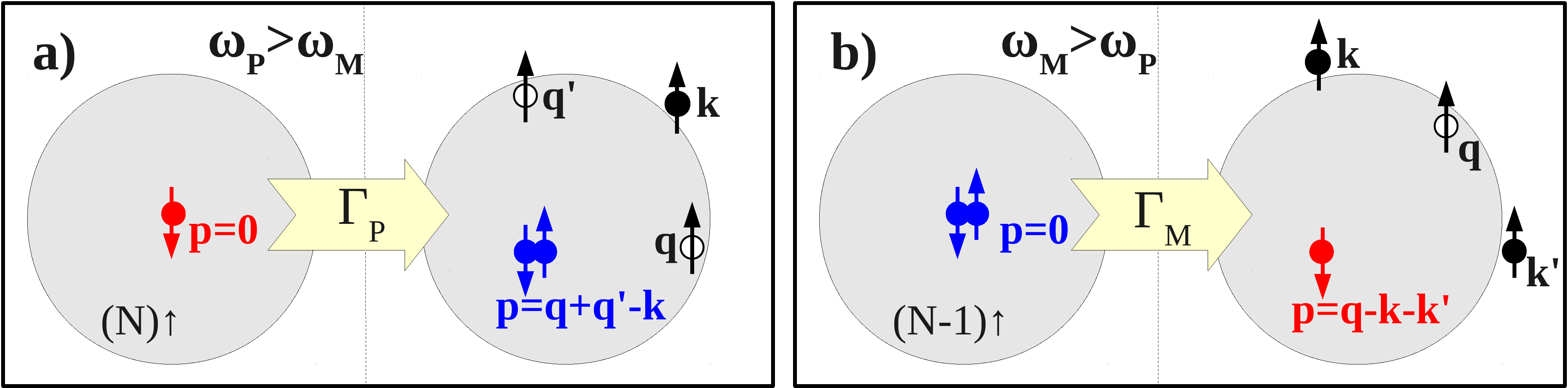}
\caption{Polaron to molecule (a) and molecule to polaron (b) decay processes in presence of a fully-polarized Fermi sea.}
\label{Kinematics}
\end{center}
\end{figure}

We consider a single $\downarrow$ impurity (which may be bosonic or fermionic)  with mass $m_\downarrow$ immersed in a FS of $\uparrow$ fermions with mass $m_\uparrow$ and density $n_\uparrow=k_F^3/6\pi^2$.
The interaction between the $\downarrow$ and $\uparrow$ particles is short-ranged and  described by  the scattering length $a$, whereas the interactions between identical fermions can be ignored.
Various approaches locate the polaron-molecule transition at the critical coupling $1/k_Fa_c \sim 0.9$ for $m_\uparrow=m_\downarrow$~\cite{Prokofev, Combescot2, Mora, Punk}. 
In the strong coupling regime $1/k_Fa>1/k_Fa_c$, a polaron with energy $\omega_P$ (we take $\hbar=1$) is unstable and will decay into a molecule with energy $\omega_M$ by removing a particle from the FS. Due to the conservation of both momentum and energy, the decay must involve the creation of an additional particle-hole pair~\cite{Prokofev}.
The leading decay channel is therefore the three-body process illustrated in Fig.\ \ref{Kinematics} (a).
Likewise for $1/k_Fa<1/k_Fa_c$, the molecule decays into a polaron by adding a fermion to the FS and creating an additional particle-hole excitation.
This is again a three-body process as illustrated in Fig.\ \ref{Kinematics} (b).

A polaron with momentum ${\mathbf{p}}$ is described by the imaginary time propagator 
$G_\downarrow({\mathbf{p}},\tau)=-\langle T_\tau[\hat{a}_{\mathbf{p}\downarrow}(\tau)\hat{a}_{\mathbf{p}\downarrow}^\dagger(0)]\rangle$. 
Here, $\hat{a}_{\mathbf{p}\downarrow}^\dagger$ creates a $\downarrow$ particle with momentum ${\mathbf{p}}$, $T_\tau$ denotes time ordering with respect 
to  $\tau\in[0;1/T[$ with $T$ the temperature (we take $k_B=1$) and $\langle\ldots\rangle$ the thermal average.
 Its Fourier transform reads
\begin{equation}
G_\downarrow({\mathbf{p}},z)^{-1}=G_\downarrow^0({\mathbf{p}},z)^{-1}-\Sigma_P({\mathbf p},z)
\end{equation}
where $G_\downarrow^0({\mathbf{p}},z)=1/(z-\xi_{p\downarrow})$,  $\xi_{p\sigma}=p^2/2m_\sigma-\mu_\sigma$, and $\mu_\sigma$ is the chemical potential for $\sigma=\uparrow,\downarrow$ species.  We take the ideal gas value $\mu_\uparrow=\epsilon_F=k_F^2/2m_\uparrow$
whereas $\mu_\downarrow$ plays no role as there is no macroscopic population 
of the $\downarrow$ states.
The self energy $\Sigma_P$ describes the effects of interactions.  
To obtain a systematic description of the  decay,  we expand the self energy as 
\begin{equation}
\Sigma_P({\mathbf p},z)=\Sigma_P^{(1)}({\mathbf p},z)+\Sigma_P^{(2)}({\mathbf p},z)+\ldots\ .
\label{Expansion}
\end{equation}
Here, $\Sigma^{(n)}_P$ involves processes with  $n$ holes in the FS.

\begin{figure}
\begin{center}
\includegraphics[width=\columnwidth]{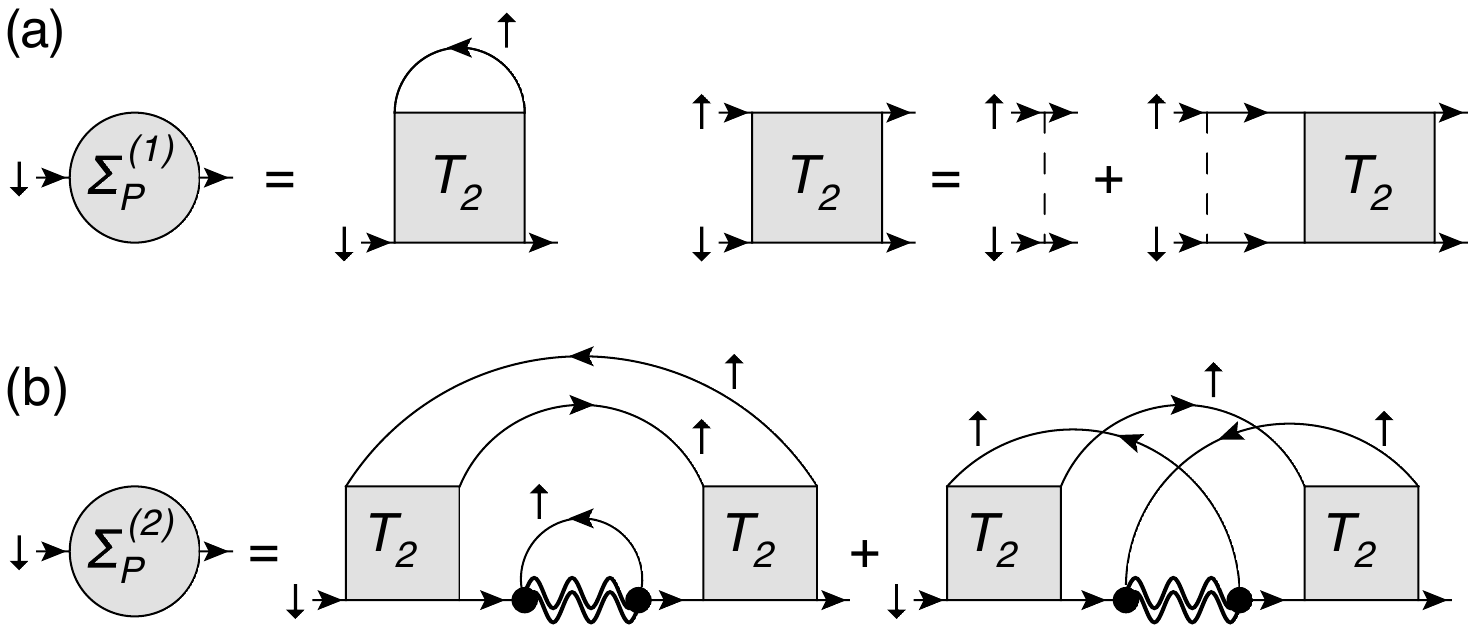}
\caption{(a) The one-hole polaron self energy $\Sigma_P^{(1)}$. (b) The two-hole polaron self energy $\Sigma_P^{(2)}$. Thin solid lines indicate $G_\sigma^0$, 
dashed lines the interaction, double wavy lines the molecule $D$, $\bullet$ the molecule-atom coupling strength $g$, and squares the scattering matrix $T_2$. }
\label{Feynman}
\end{center}
\end{figure}

 The term $\Sigma_P^{(1)}$  shown in Fig.\ \ref{Feynman} (a) corresponds to the commonly used ladder approximation.  The energy of a zero-momentum polaron obtained from
 \begin{equation}
 \omega_P=\Sigma_P^{(1)}({\mathbf p}=0,\omega_P+i0_+)
 \label{PolaronEn}
\end{equation}
(we drop the infinitesimal $i0_+$ in the following) is essentially indistinguishable from the Monte-Carlo result~\cite{Chevy,Alessio,Massignan,Pilati,Prokofev}.
However, (\ref{PolaronEn}) yields a real energy for any coupling strength. To obtain the damping in the metastable regime it is therefore necessary to include processes involving creation of at least two holes.

The Feynman diagrams for $\Sigma_P^{(2)}$ corresponding to the decay process in Fig.\ \ref{Kinematics} (a) are 
given in Fig.\ \ref{Feynman} (b). 
They describe the  scattering of a ${\mathbf{p}}=0$ polaron  creating two holes with momenta ${\mathbf{q}}$ and ${\mathbf{q}}'$,  and leaving one fermion above the FS with 
momentum ${\mathbf{k}}$.  The remaining $\uparrow\downarrow$ pair can then form a molecule  with momentum ${\mathbf{q}}+{\mathbf{q}}'-{\mathbf{k}}$.

\begin{figure}
\begin{center}
\includegraphics[width=\columnwidth]{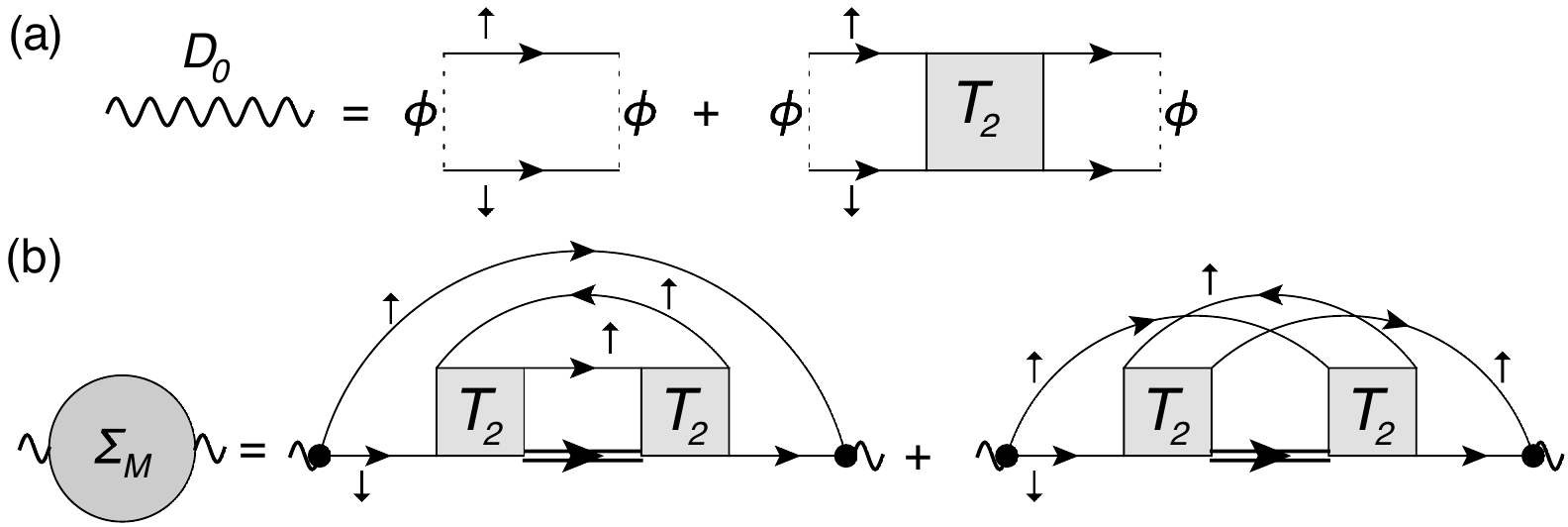}
\caption{(a) The molecule propagator in the ladder approximation. (b) Diagrams corresponding to the leading molecule decay channel. Wavy lines denote the molecule $D^0$ in the ladder approximation, dotted lines with $\phi$
the operator  $\hat b$ or $\hat b^\dagger$, and
double lines the polaron propagator ($G_\downarrow - G_\downarrow^0$).}
\label{FeynmanMol}
\end{center}
\end{figure}

The propagator of the open-channel,  Feshbach molecule is $D({\mathbf{q}},\tau)=-\langle T_\tau[\hat{b}_{\mathbf{p}}(\tau)\hat{b}_{\mathbf{p}}^\dagger(0)]\rangle$ with 
$\hat{b}_{\mathbf{p}}^\dagger=\int d^3\check q\phi_q\hat{a}_{{\mathbf{p}}/2+{\mathbf{q}}\downarrow}^\dagger\hat{a}_{{\mathbf{p}}/2-{\mathbf{q}}\uparrow}^\dagger$ 
and  $d^3\check q= d^3q/(2\pi)^3$.
Ignoring finite range effects, its wave function in vacuum reads  
 $\phi_q=\sqrt{8\pi a^3}/(1+q^2a^2)$. The Fourier transform of the molecule propagator can be written as
\begin{equation}
D({\mathbf{p}},z)^{-1}=D_0({\mathbf{p}},z)^{-1}-\Sigma_M({\mathbf{p}},z)
\end{equation}
where $\Sigma_M$ is the molecule self energy and  
\begin{equation}
D_0({\mathbf{p}},z)=\int d^3\check q \phi_q^2\frac{1-f(\xi_{{\mathbf p}-{\mathbf q}\uparrow})}{z-\xi_{{\mathbf p}-{\mathbf q}\uparrow}
-\xi_{q\downarrow}}+\frac{T_2({\mathbf{p}},z)}{g({\mathbf{p}},z)^2}
\end{equation}
describes the propagation of a $\uparrow\downarrow$ pair in the ladder approximation. It is illustrated in Fig.\ \ref{FeynmanMol} (a). 
Here $f(x)=[\exp (x/T)+1]^{-1}$ is the Fermi function and 
\begin{equation}
\frac 1 {g({\mathbf{p}},z)}=\int d^3\check q \phi_q\frac{1-f(\xi_{{\mathbf p}-{\mathbf q}\uparrow})}{z-\xi_{{\mathbf p}-{\mathbf q}\uparrow}
-\xi_{q\downarrow}}.
\label{moleculeAtomCoupling}
\end{equation}
is the atom-molecule coupling. The object entering the diagrammatic analysis is $g({\mathbf{p}},z)^2D({\mathbf{p}},z)$, which describes the scattering of $\uparrow\downarrow$ atoms mediated by the Feshbach molecule~\cite{BruunPethick}.
In the following we use the vacuum limit of  (\ref{moleculeAtomCoupling}), $g=-\sqrt{2\pi/m_r^2a}$ with $m_r=m_\uparrow m_\downarrow/(m_\downarrow+m_\uparrow)$. 

Using the fact that there is no macroscopic population of the $\downarrow$ state and that the $\uparrow$ fermions form an ideal gas,  the frequency sums in the two diagrams can be performed analytically by contour integration. This amounts to evaluating frequencies at the on-shell $\uparrow$ energies.
The polaron decay rate is given by  \mbox{$\Gamma_P=-{\rm{Im}}\Sigma^{(2)}_P(0,\omega_P)$}.
 When $0<\Delta\omega=\omega_P-\omega_M\ll \epsilon_F$, the 
$G^0_\downarrow$'s and $T_2$'s in Fig.\ \ref{Feynman} (b) are off resonant and therefore real. The only contribution to the imaginary part giving rise to damping comes from the Feshbach molecule, which in the vicinity of its pole can be described by 
\begin{equation}
D({\mathbf{p}},\omega)\simeq \frac{Z_M}{\omega-\omega_M-p^2/2m_M^*}.
\label{Pole}
\end{equation}
Here,  $Z_M$ and $m_M^*$  are the molecule residue and effective mass respectively. 
The contribution to $\Gamma_P$ from the two diagrams in Fig.\ \ref{Feynman} (b)  at $T=0$ can after some algebra  be written in the symmetric form
\begin{gather}
\Gamma_P=\frac{g^2 Z_M}{2}\int d^3\check q d^3\check k d^3\check q'
\left[F({\mathbf{q}},{\mathbf{k}},\omega_P)-F({\mathbf{q}}',{\mathbf{k}},\omega_P)\right]^2
\nonumber\\
\times\delta\left(\Delta\omega+\xi_{q\uparrow}+\xi_{q'\uparrow}-\xi_{k\uparrow}-({\mathbf{q}}+{\mathbf{q}}'-{\mathbf{k}})^2/2m_M^*\right).
\label{PolDecayRate}
\end{gather}
We have defined $F({\mathbf{q}},{\mathbf{k}},\omega)=T_2({\mathbf{q}},\omega+\xi_{q\uparrow})G_\downarrow^0({\mathbf{q}}-{\mathbf{k}},\omega+\xi_{q\uparrow}-\xi_{k\uparrow})$.   Here and in the following, we take $q,q'<k_F$ and $k>k_F$. The summation over two holes and one particle in (\ref{PolDecayRate}) can be interpreted as 
a reorganization of the FS to contain one less particle.

For $\Delta\omega\ll \epsilon_F$, the holes and particles involved in the scattering process illustrated in Fig.\ \ref{Kinematics}(a) are located around the Fermi surface, i.e.\ $q\simeq k\simeq k'\simeq k_F$. Also, the molecule momentum is small since $p^2/2m_M^*<\Delta\omega$. 
 Using this and ignoring for the time being  the matrix element  in (\ref{PolDecayRate}),  we get the integral 
 \begin{gather}
 \int d^3\check p\int_0^\infty d\xi\int_{-\epsilon_F}^0 d\xi'd\xi''\delta\left(\Delta\omega+\xi'+\xi''-\xi-\frac{p^2}{2m_M^*}\right)\nonumber\\
=\frac{ 2^{5/2}}{105\pi^2}(m_M^*)^{3/2}\Delta\omega^{7/2}.
\label{Integral}
\end{gather}
Thus, the available  phase space of the three-body process gives a factor $\Delta\omega^{7/2}$ for  the decay rate: the $\delta$-function removes one energy integral in (\ref{Integral}), a factor $\Delta\omega$ comes from each of the two remaining degenerate energies, and a factor $\Delta\omega^{3/2}$ comes from the molecule.
Let us now focus on the matrix element $F({\mathbf{q}},{\mathbf{k}},\omega_P)-F({\mathbf{q}}',{\mathbf{k}},\omega_P)$ in (\ref{PolDecayRate}).
The momenta ${\mathbf{q}}$, ${\mathbf{q}}'$, and ${\mathbf{k}}$ form an equilateral triangle for $\Delta\omega=0$ and
the matrix element  vanishes, since $F({\mathbf{q}},{\mathbf{k}},\omega)$ only depends on the angle between ${\mathbf{q}}$ and ${\mathbf{k}}$.
 For $\Delta\omega\ll \epsilon_F$, we can  expand the matrix element in the deviations of the triangle formed by ${\mathbf{q}}$, ${\mathbf{q}}'$, and ${\mathbf{k}}$ away from the equilateral
shape. Using this in (\ref{PolDecayRate}) yields
\begin{equation}
\Gamma_P\sim Z_Mk_Fa\left(\frac{\Delta\omega}{\epsilon_F}\right)^{9/2}\epsilon_F.\label{LifeTime}
\end{equation}
Thus, the Fermi antisymmetry when swapping the hole momenta 
${\mathbf q}$ and ${\mathbf q}'$ gives an additional factor $\Delta\omega$ to the decay rate. The resulting 
polaron lifetime $1/2\Gamma_P$  diverges much faster than $1/\Delta\omega$ close to the transition point $k_Fa_c$. 

From (\ref{LifeTime}) it follows that the matrix element between equal energy  states which contain a molecule or a polaron vanishes at the transition point. 
This means that there is no avoided crossing and that  the transition between the two states is of first order, as was suggested previously~\cite{Prokofev}.

Consider now the regime $\Delta\omega<0$ ($1/k_Fa<1/k_Fa_c$). 
Here the  molecule  decays into a  polaron  via the three-body process  illustrated in Fig.\ \ref{Kinematics}(b).
The corresponding Feynman diagrams are shown in \ref{FeynmanMol}(b). 
After the molecule has split, the impurity scatters with the FS leaving one hole with momentum ${\mathbf{q}}$ and two particles with momenta ${\mathbf{k}}$ and ${\mathbf{k}}'$.
 The impurity can then form a polaron with momentum ${\mathbf{p}}={\mathbf{q}}-{\mathbf{k}}-{\mathbf{k}}'$.
The decay rate for a zero momentum  molecule is given by \mbox{$\Gamma_M=-{\rm{Im}}\Sigma_M(0,\omega_M)$}, and calculations like the ones for the polaron  show that for $T=0$ and $|\Delta\omega|\ll \epsilon_F$ the two diagrams give 
\begin{gather}
\Gamma_M=\frac{g^2 Z_P}{2}\int d^3\check k d^3\check k' d^3\check q
\left[C({\mathbf{q}},{\mathbf{k}},\omega_M)-C({\mathbf{q}},{\mathbf{k}}',\omega_M)\right]^2
\nonumber\\
\times\delta\left(|\Delta\omega|+\xi_{q\uparrow}-\xi_{k\uparrow}-\xi_{k'\uparrow}-({\mathbf{q}}-{\mathbf{k}}-{\mathbf{k}}')^2/2m_P^*\right).
\label{MolDecayRate}
\end{gather}
Here, $C({\mathbf{q}},{\mathbf{k}},\omega)=G_\downarrow^0({\mathbf{k}},\omega-\xi_{k\uparrow})T_2({\mathbf{k}}-{\mathbf{q}},\omega-\xi_{k\uparrow}+\xi_{q\uparrow})$
and $Z_P$ is the polaron residue. 
\begin{figure}
\begin{center}
\includegraphics[angle=-90,width=\columnwidth]{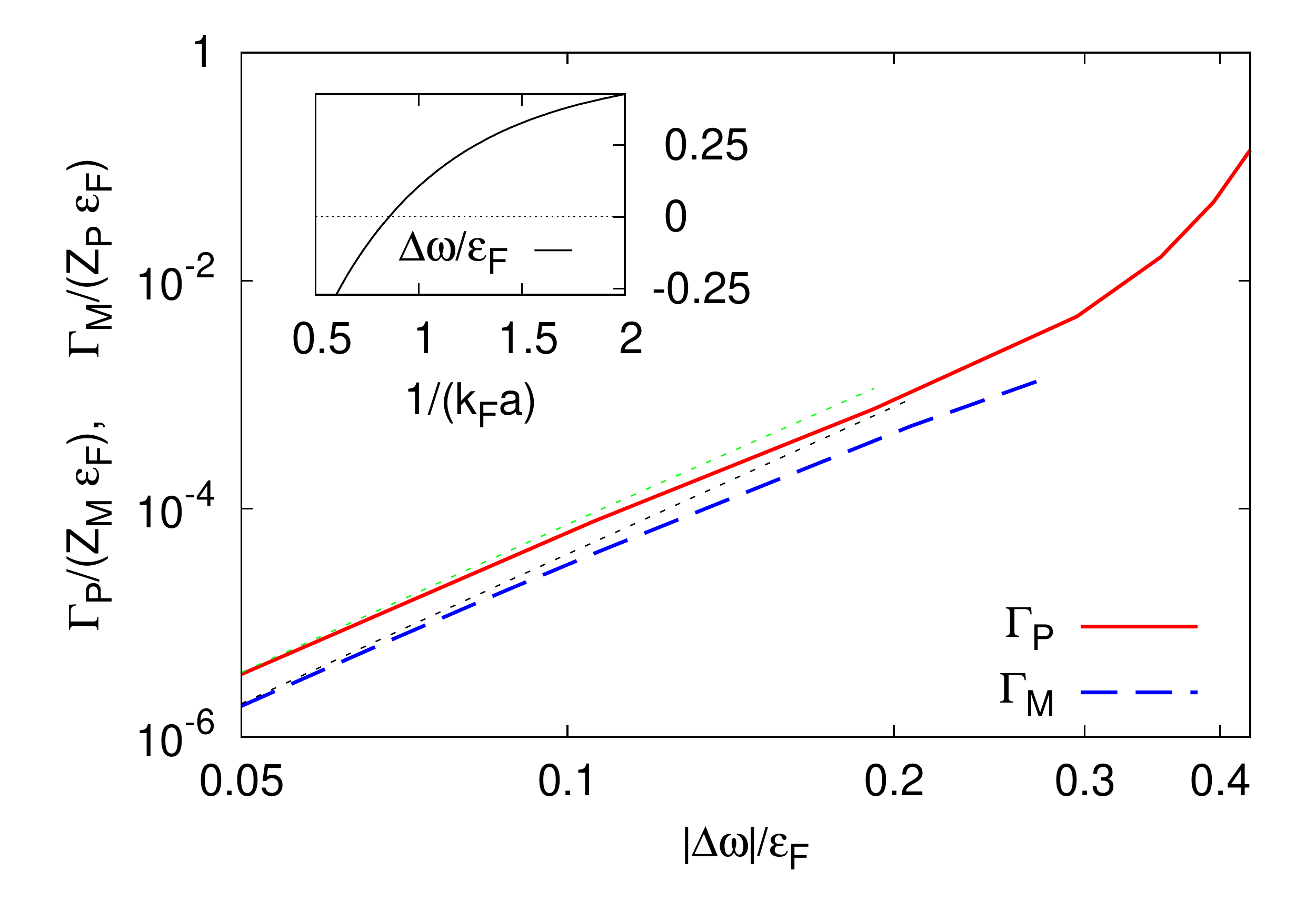}
\caption{Polaron$\rightarrow$molecule and molecule$\rightarrow$polaron decay rates $\Gamma_P$ and $\Gamma_M$ as a function of the energy difference $|\Delta\omega|$ for $m_\uparrow=m_\downarrow$ and $T=0$. The thin dashed lines are the predicted asymptotic behaviours, $\Gamma_P\propto(m_M^*)^{3/2}\Delta\omega^{9/2}$ and $\Gamma_M\propto(m_P^*)^{3/2}\Delta\omega^{9/2}$. Inset: $\Delta\omega$ as a function of $1/k_Fa$.}
\label{numerics}
\end{center}
\end{figure}
Like for the polaron decay, one can derive from (\ref{MolDecayRate}) that 
\begin{equation}
\Gamma_M\sim Z_Pk_Fa\left(\frac{|\Delta\omega|}{\epsilon_F}\right)^{9/2}\epsilon_F
\label{MolLifeTime}
\end{equation}
for $|\Delta\omega|\ll\epsilon_F$.
 Again, this means that the molecule is long lived close to the cross-over.

 In Fig.~(\ref{numerics}) we show calculations of the decay rates obtained from Eqs.~(\ref{PolDecayRate}) and (\ref{MolDecayRate}) for the $m_\uparrow=m_\downarrow$ case. In the numerics we have used
 \begin{equation}
T_2({\mathbf{p}},\omega)=
 \frac{2\pi a/m_r}{1-\sqrt{2m_ra^2(p^2/2m_M-\omega-\epsilon_F+g_{3} n_\uparrow)}},
\end{equation}
which describes molecules with energy
 \begin{equation}
\omega_M=-\hbar^2/(2m_r a^2)-\epsilon_F+g_{3}n_\uparrow.
\label{molEnergy}
\end{equation}
Here $g_{3}=2\pi\hbar^2a_3/m_3$, $m_3=m_\uparrow (m_\uparrow+m_\downarrow)(2m_\uparrow+m_\downarrow)$, and $a_3=1.18a$ is the molecule-atom scattering length~\cite{Skorniakov,Brodsky}. Equation (\ref{molEnergy}) is exact in the limit $1/k_Fa\rightarrow\infty$, and accurate all the way to the critical coupling $1/k_Fa_c$ for the case $m_\uparrow=m_\downarrow$~\cite{Prokofev}. The polaron energy is obtained from (\ref{PolaronEn}), while  $m^*_P$ and $m^*_M$ are taken from Ref.~\cite{Prokofev}.
 The numerical results shown  in Fig.~\ref{numerics} match nicely   with (\ref{LifeTime}) and (\ref{MolLifeTime}) for $|\Delta\omega|\ll \epsilon_F$. In the fits, we have taken into account the factors $(m_P^*)^{3/2}$ and $(m_M^*)^{3/2}$ coming from the $\delta$-functions [see (\ref{Integral})]. 
  For larger $|\Delta\omega|$, there is as expected some discrepancy between the numerical results and  (\ref{LifeTime}) and (\ref{MolLifeTime}) since the momenta involved in the scattering processes are no longer close to the Fermi surface.

Our analysis shows that the decay of polarons and molecules will happen on timescales of order $10-100$ms. 
To observe the decay, one could for instance sweep the scattering length across $a_c$. 
  Zener type of arguments predict that the polarons/molecules will survive as metastable states after the sweep~\cite{Zener}. 
Radio-frequency spectroscopy~\cite{Schirotzek} or optical probes can then  be used to map the exponential decrease in time of the polaron/molecule population.
 Alternatively, one can extract the effective mass from collective mode frequencies~\cite{Bruun}. This technique has been successfully applied at unitarity~\cite{Nascimbene}. By measuring the change in the frequency as a function of time, one may  extract the decay rate of the metastable state.

The polaron can decay to deeply bound closed channel molecules via three-body processes analogous to the ones considered here for the decay into the weakly bound 
Feshbach molecule.
 This decay is however strongly suppressed because the deeply bound molecules are spatially much smaller than the Feshbach molecule, whose size is $\sim a$.
Since the decay requires two $\uparrow$ fermions to be at distance of order of the molecule size, the Pauli principle suppresses decay into deep molecular states much more efficiently than into the larger Feshbach state~\cite{Petrov}. The Feshbach molecule furthermore has a large component in the open channel for broad resonances, which justifies our use of a single channel theory in the present paper. Contrary to this, the deeply bound molecules are mostly in the closed channel where the fermions are in different spin states. This should suppress the decay even further as it must involve spin flips.  Indeed, no decay to deeply bound states has been observed for a 
strongly polarized Fermi gas~\cite{Schirotzek}, and  very long lifetimes of order tens of seconds have been measured for 
balanced two component Fermi gases~\cite{Jochim}. Contrary to this, our results show that the decay into the weakly bound Feshbach molecule is in the 
10-100ms range, confirming that it is the dominant loss process. 

One can include more interaction effects in the decay for instance by replacing $G^0_\downarrow$ with $G_\downarrow$ in the matrix elements
 or by introducing additional scattering events.
This will change the quantitative value of the matrix elements for the scattering processes but will not change the $\Gamma\sim |\Delta\omega|^{9/2}$ scaling as this comes from the combined effects of kinematics and Fermi statistics; it can in fact be obtained from a Golden Rule calculation.

For $m_\uparrow=m_\downarrow$ and $T=0$, Monte-Carlo calculations
 predict  at small impurity densities a phase separated state before polarons may decay into molecules~\cite{Pilati}.
We expect however that non-zero temperatures will stabilize the polaron state against phase separation due to the entropy of mixing. 
Moreover, the boundary between mixed and separated states depends on the ratio $m_\downarrow / m_\uparrow$~\cite{Pao}. By appropriately selecting the atomic species, one may bring the critical crossing point back to a physically observable region.
Alternatively, one could use bosonic impurities, for which stability against phase separation has recently been predicted~\cite{Marchetti}. It would therefore be interesting to extend experiments with large population imbalances to mass-imbalanced fermionic mixtures, or to Bose-Fermi mixtures.

In conclusion, we have considered the leading three-body processes involved in the polaron-molecule coupling.
 Our analysis shows that the coupling is strongly suppressed due to a combination of phase-space and Pauli blocking effects, vanishing as a power law close to the transition.
This yields very long lifetimes of the metastable states, and implies a first order transition between the two ground states.
Our results suggest new directions for experiments with polarized atomic gases.

\acknowledgments
We acknowledge useful discussions with C.\ J.\ Pethick. P.M.\ acknowledges ESF/MEC project FERMIX (FIS2007-29996-E), Spanish MEC projects FIS2008-01236, FIS2008-00784, QOIT (Consolider Ingenio 2010) and Catalan project 2009-SGR-985.

\end{document}